\documentclass[noshowpacs,amsmath,
twocolumn,
superscriptaddress,
8pt,
aps,prb]{revtex4-1}
\usepackage{setspace}
\usepackage{amsmath}
\usepackage{graphicx}
\usepackage[nearskip,margin = 0pt]{subfig}

\usepackage{verbatim}
\usepackage{amsfonts}
\usepackage{amssymb}
\usepackage{epstopdf} 
\usepackage{xcolor}
\usepackage{verbatim}
\usepackage{upgreek}
\usepackage{ragged2e}
\DeclareGraphicsExtensions{.pdf,.eps,.png,.jpg,.mps} 

\begin{document}
\title{A squeezed quantum microcomb on a chip}
\author{Zijiao Yang$^{1,2,*}$, Mandana Jahanbozorgi$^{1,*}$, Dongin Jeong$^{3}$, Shuman Sun$^{1}$, Olivier Pfister$^{2}$, Hansuek Lee$^{3,4}$ and Xu Yi$^{1,2,\dagger}$\\
\vspace{3pt}
$^1$Department of Electrical and Computer Engineering, University of Virginia, Charlottesville, Virginia 22904, USA.\\
$^2$Department of Physics, University of Virginia, Charlottesville, Virginia 22904, USA.\\
$^3$Graduate School of Nanoscience and Technology, Korea Advanced Institute of Science and Technology (KAIST), Daejeon 34141, South Korea.\\
$^4$Department of Physics, Korea Advanced Institute of Science and Technology (KAIST), Daejeon 34141, South Korea.\\
$^{\ast}$These authors contributed equally to this work.\\
$^{\dagger}$Corresponding author: yi@virginia.edu}

 
\date{\today}

\maketitle

    \noindent {\bf The optical microresonator-based frequency comb (microcomb) provides a versatile platform for nonlinear physics studies\cite{kippenberg2018dissipative} and has wide applications ranging from metrology\cite{spencer2018optical} to spectroscopy\cite{suh2016microresonator}. Deterministic quantum regime is an unexplored aspect of microcombs, in which unconditional entanglements among hundreds of equidistant frequency modes can serve as critical ingredients to scalable universal quantum computing\cite{menicucci2008one,yokoyama2013ultra,chen2014experimental,asavanant2019generation,larsen2019deterministic, pfister2019continuous} and quantum networking\cite{roslund2014wavelength,guo2020distributed}. Here, we demonstrate a deterministic quantum microcomb in a silica microresonator on a silicon chip. 40 continuous-variable quantum modes, in the form of 20 simultaneously two-mode squeezed comb pairs, are observed within 1 THz optical span at telecommunication wavelengths. A maximum raw squeezing of 1.6 dB is attained. A high-resolution spectroscopy measurement is developed to characterize the frequency equidistance of quantum microcombs. Our demonstration offers the possibility to leverage deterministically generated, frequency multiplexed quantum states and integrated photonics to open up new avenues in fields of spectroscopy\cite{shi2020entanglement}, quantum metrology\cite{anisimov2010quantum}, and scalable quantum information processing\cite{weedbrook2012gaussian}. }

\noindent

Optical microresonators employ Kerr nonlinearity \cite{kippenberg2004kerr} to provide broadband parametric gain through four-wave mixing (FWM) among cavity resonance modes, where pairs of pump photons can be annihilated to generate signal and idler photons at lower and higher frequencies. Above the Kerr parametric oscillation threshold, microcombs behave classically as Kerr nonlinearity offsets both cavity dispersion and loss to create dissipative Kerr solitons\cite{kippenberg2018dissipative}. Below the threshold, the microcomb is dominated by the quantum correlation of the co-generated photon pairs\cite{chembo2016quantum}. Quantum microcomb can provide hundreds of frequency multiplexed quantum channels from a single microresonator\cite{kues2019quantum}. Access to individual quantum channels is possible through the wavelength-division-multiplexing technology thanks to microcombs' large free-spectral-ranges (FSRs). When combined with integrated photonic circuits, quantum microcombs have the potential to revolutionize photonic quantum information processing.

However, so far, experiments of quantum microcombs have been limited to the probabilistic regime\cite{reimer2016generation,kues2017chip,imany201850,kues2019quantum}, where entanglements are measured between randomly emitted photon pairs with postselection by coincident detection. The photon coincidence rate suffers from exponential decrease with the increase of photon number in a quantum state\cite{kues2017chip}. Quantum architectures built upon probabilistic quantum states are not scalable without quantum memory\cite{lvovsky2009optical}. In addition, no high resolution method is reported to verify the frequency equidistance, a prerequisite of frequency combs, for the probabilistic quantum microcombs. A quantum microcomb in the deterministic regime will be a significant step forward towards the scalable quantum architecture on photonic chips.

One approach to construct deterministic quantum microcombs is to leverage two-mode squeezing and create unconditional entanglement between the optical fields in optical frequency combs\cite{pysher2011parallel,chen2014experimental,wu2020quantum}. Squeezed light, with quantum uncertainty below than that of the vacuum field, has broad applications in science and technology, ranging from enhancing the gravitational wave detection sensitivity in LIGO\cite{aasi2013enhanced}, Gaussian boson sampling\cite{Zhongeabe8770}, to generating large-scale cluster states for quantum computing\cite{yokoyama2013ultra,chen2014experimental,asavanant2019generation,larsen2019deterministic,pfister2019continuous} and networking\cite{roslund2014wavelength}. The field entanglement created by two-mode squeezing is unconditional, nonclassical and does not rely on postselection. As a result, the number of squeezed quantum modes (qumodes) in a quantum state can be deterministically scaled up \cite{chen2014experimental,roslund2014wavelength,yokoyama2013ultra,asavanant2019generation,larsen2019deterministic}, which provides a scalable physical platform for continuous-variable-based quantum computing\cite{pfister2019continuous}. Squeezed quantum microcombs, when combined with integrated photonic circuits, Gaussian and non-Gaussian measurements, can serve as simple and compact building bricks for universal quantum computing\cite{zhu2021Hypercubic}, entanglement-assisted spectroscopy\cite{shi2020entanglement}, and quantum networking for distributed quantum sensing\cite{guo2020distributed}. While generation of one or two squeezed frequency qumodes \cite{kanter2002squeezing,lenzini2018integrated,otterpohl2019squeezed,kashiwazaki2020continuous,zhao2020near,vaidya2020broadband} and their detection\cite{tasker2020silicon} in miniaturized platforms have been shown recently, a squeezed microcomb has not been reported yet.

\begin{figure*}[!bht]
\captionsetup{singlelinecheck=off, justification = RaggedRight}
\includegraphics[width=17cm]{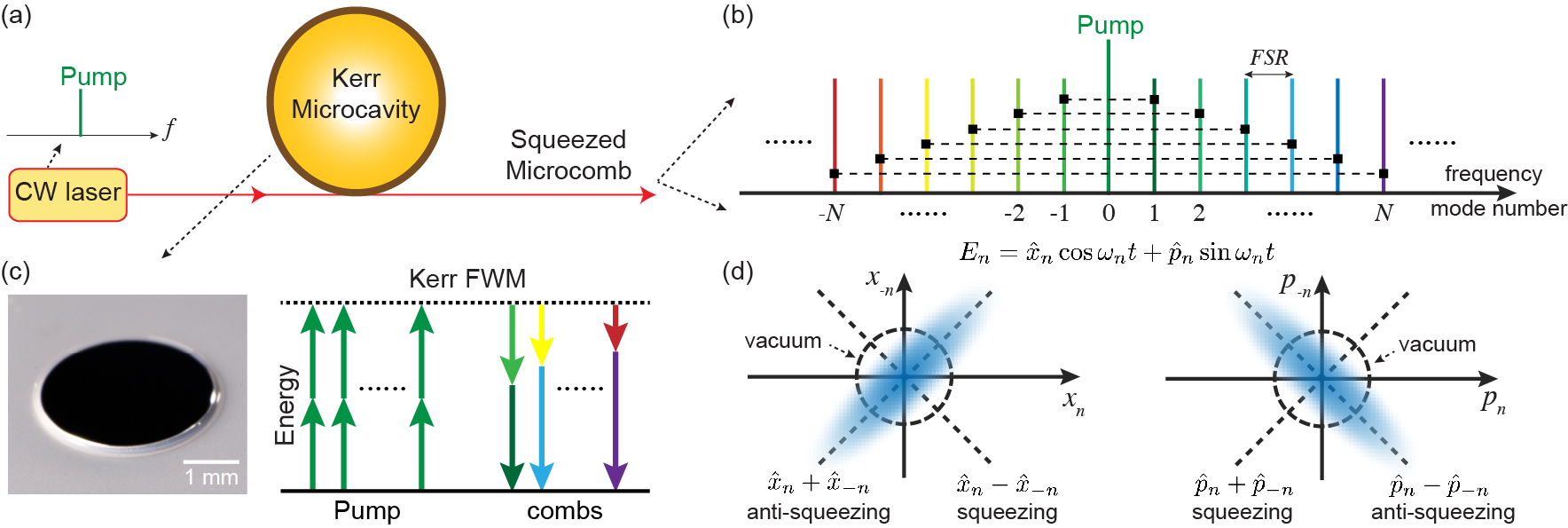}
\caption{{\bf Generation of deterministic, two-mode squeezed quantum microcombs on a chip.} {\bf (a)} A continuous-wave (cw) pump laser is coupled to a microresonator, which has thousands of longitude resonance modes with their frequencies separated by the resonator free-spectral-range (FSR), as shown in panel {\bf (b)}. {\bf (c)} The $\chi^{(3)}$ Kerr nonlinearity in the microresonator creates broadband parametric gain as the pump photon pairs (green) can be converted into signal and idler photons at lower and higher frequency modes. This non-classical correlation creates two-mode vacuum squeezing and thus unconditional EPR entanglement of the optical quadrature fields between frequency modes $n$ and $-n$, which are connected by dashed black lines in the optical spectrum in panel (b). Also shown is the image of a silica microresonator on a silicon chip used in this work. {\bf (d)} Conceptual illustration of the two-mode squeezing wavefunctions in position (left) and momentum (right) basis, where ($\hat{x}_n$ - $\hat{x}_{-n}$) and ($\hat{p}_n$ + $\hat{p}_{-n}$) have uncertainty level below the vacuum fluctuation (dashed circle). The electrical field of the $n$-th optical mode is ${E}_n= \hat{x}_n \cos{\omega_n t } + \hat{p}_n \sin{\omega_n t }$, where $\hat{x}_n$ and $\hat{p}_n$ are the in-phase and out-of-phase quadrature amplitudes of the mode at frequency $\omega_n$.}
\label{fig:concept}
\end{figure*}

In this work, we demonstrate a deterministic, two-mode-squeezed quantum frequency comb in a silica microresonator on a silicon chip (Fig.\ref{fig:concept}). The Kerr parametric process generates unconditional Einstein-Podolsky-Rosen (EPR) entanglement, i.e., two-mode squeezing, between the optical quadrature fields of the qumode pairs in the microresonator (Fig. \ref{fig:concept}). 40 frequency multiplexed qumodes, in the form of 20 two-mode squeezed comb pairs, are measured at telecommunication wavelengths. The two-mode squeezing is verified by measuring the noise variance by means of balanced homodyne detection. Maximum raw squeezing of 1.6 dB and maximum anti-squeezing of 6.5 dB are attained. A corresponding 3.1 dB squeezing at the output waveguide can be inferred after correcting system losses. A qumode spectroscopy measurement is developed to characterize the frequencies of qumodes with a resolution of 5 MHz in 1 THz optical span. The fundamental resolution limit of our spectroscopy method is set by the microcavity resonance linewidth. In our experiments, the number of accessible qumodes is limited by the 1 THz optical span of the local oscillators. The number of qumodes can exceed 1,000 when the resonator dispersion is properly engineered.

\begin{figure*}[!bht]
\captionsetup{singlelinecheck=off, justification = RaggedRight}
\includegraphics[width=17cm]{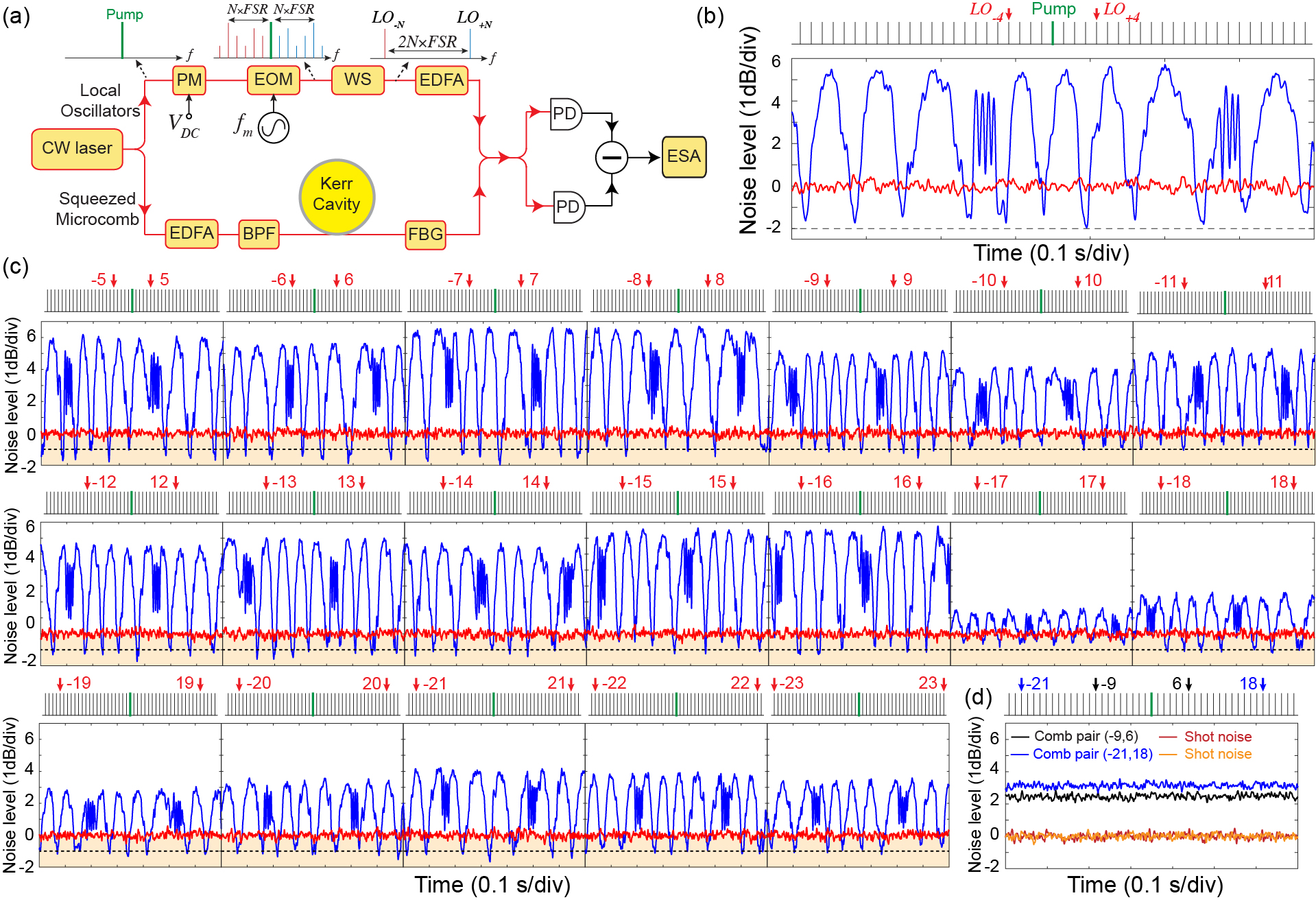}
\caption{{\bf Two-mode squeezing measurement of 20 comb pairs (40 qumodes) from the microresonator.} {\bf(a)} Simplified experimental schematic. A continuous-wave (cw) laser is split to pump the resonator and drive the local oscillators (LOs). The LOs are derived from an electro-optics modulation (EOM) frequency comb, with a comb spacing (modulation frequency) of $f_m$. A line-by-line waveshaper (WS) is used to select a pair of comb lines as the bichromatic local oscillators. The  phase of the LOs can be tuned by a phase modulator (PM). The LOs and the squeezed microcombs are combined by a 50/50 coupler and are detected on balanced photodiodes (PDs). The noise level is characterized on an electrical spectrum analyzer (ESA). In the squeezed microcomb path, a fiber Bragg grating (FBG) filter is used to block the strong pump light. Erbium-doped fiber amplifiers (EDFAs) and bandpass filter (BPF) are also shown in the figure. {\bf(b)} Representative quadrature noise variance (blue) relative to shot noise (red) as a function of time for qumodes -4 and 4 (indicated with red arrows). The phase of the LOs ramps periodically with time (slow up-scan, and fast down-scan). 1.6 dB squeezing and 5.5 dB anti-squeezing are directly observed. A dashed black line indicates 2 dB below shot noise level. {\bf(c)} Quadrature noise variance (blue) relative to shot noise (red) of all 40 qumodes. The qumodes measured are marked by the red arrows. The regime below shot noise limit is colored in orange, and a dashed black line indicates 1 dB below shot noise level. (d) Entanglement check: noise variances show no quantum correlation between uncorrelated comb pairs for qumodes (-9, 6) and (-21, 18). All measurements are taken at 2.7 MHz frequency, 100 kHz resolution bandwidth, and 100 Hz video bandwidth.}
\label{fig:data1}
\end{figure*}

\begin{figure}[!bht]
\captionsetup{singlelinecheck=off, justification = RaggedRight}
\includegraphics[width=8cm]{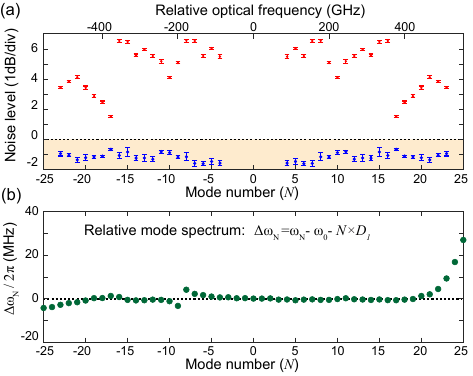}
\caption{{\bf Summary of squeezing and anti-squeezing levels and resonator mode spectrum.} {\bf(a)} Squeezing (blue) and anti-squeezing (red) levels versus mode number. The regime below the shot noise level is colored in orange. It should be noted that the noise level at qumode $N$ or $-N$ represents the two-mode noise level of comb pair $(-N, N)$. The error bars are concluded with a 95$\%$ confidence interval under t-distribution. {\bf(b)} The cold resonator mode spectrum ($\Delta \omega_N$, relative mode frequency). The degradation of squeezing/anti-squeezing level of certain qumodes is likely caused by the avoided mode crossing induced by spatial-mode-interaction in the microresonator.}
\label{fig:data2}
\end{figure}

\begin{figure*}[!bht]
\captionsetup{singlelinecheck=off, justification = RaggedRight}
\includegraphics[width=17cm]{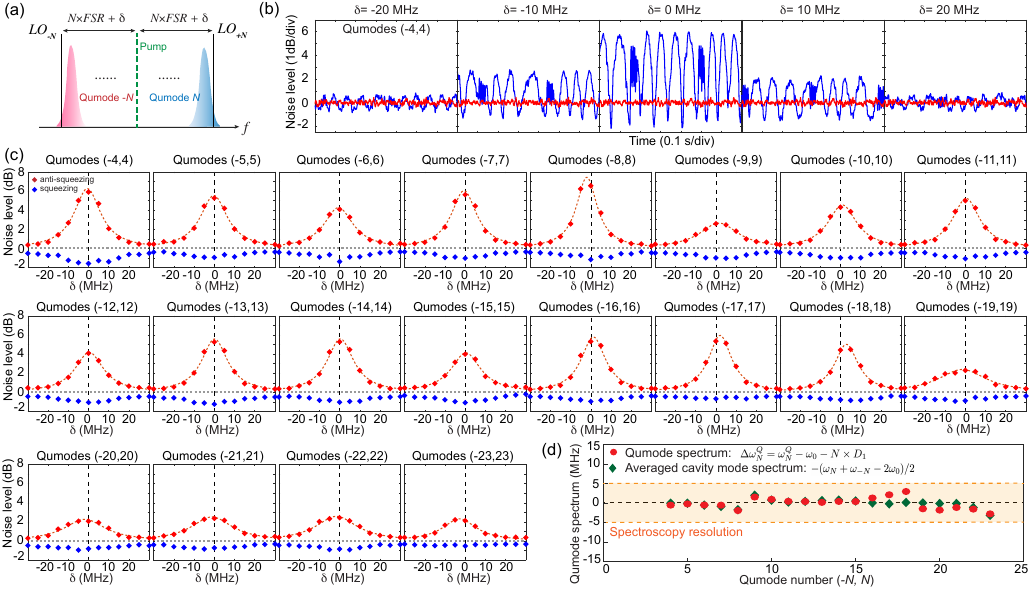}
\caption{{\bf Spectroscopy characterization of qumodes in the squeezed quantum microcomb.} {\bf (a)} Illustration of spectroscopy measurement of qumode (-$N$,$N$). The frequencies of the $N$-th LOs can be detuned by $\delta$ away from the equidistant frequencies, $\pm N \times$FSR, and the amount of squeezing and anti-squeezing are measured at each detuning point, $\delta$. {\bf (b)} Noise variance measurement of qumodes (-4,4) at detuning $\delta$ = -20, -10, 0, 10, 20 MHz. The red trace represents shot noise level. {\bf (c)} Squeezing (blue) and anti-squeezing (red) levels extracted from noise variance measurements at different detuning points ($\delta$) for all qumodes. Shot noise levels are represented by the horizontal dashed gray lines. Lorentzian fitting of the anti-squeezing spectrum  (red dash line) is used to find the qumode center frequencies. Vertical dashed black lines represent the equidistant frequencies for each qumodes. {\bf (d)} Summary of the measured relative qumode frequencies (red) from qumodes (-3,3) to (-23,23). The two-sided averaged cavity mode spectrum: -$(\Delta \omega_N$ + $\Delta \omega_{-N})/2$ is plotted in green and it agrees well with the qumode spectrum.}
\label{fig:data3}
\end{figure*}

The microresonator used in this work is a 3 mm diameter silica wedge resonator with 22 GHz free-spectral-range (FSR) on a silicon chip\cite{lee2012chemically}. The resonator's intrinsic quality factor ($Q_o$) is 79 million, and a single mode tapered fiber is used as the coupling waveguide. The resonator is overcoupled to achieve large field escape efficiency, $\eta=\kappa_{c}/(\kappa_{c} + \kappa_o)$, of 83\%, where $\kappa_0$ and $\kappa_c$ are the intracavity dissipation rate and the resonator-waveguide coupling rate. An amplified continuous-wave laser is used as the pump of the resonator, and its frequency ($\omega_p$) is phase locked to the resonance mode at 1550.5 nm through Pound–Drever–Hall (PDH) locking technique \cite{drever1983laser}. The pump power is set to 120 mW, which is slightly below the parametric threshold of 135 mW. At the through port of the resonator, a fiber Bragg grating (FBG) filter is used to filter out the pump field. In principle, the bichromatic local oscillators (LOs) for two-tone homodyne detection can be derived from a soliton microcomb\cite{kippenberg2018dissipative}. However, in the present measurement, an electro-optic modulation (EOM) frequency comb\cite{carlson2018ultrafast} is used, where tens of modulation sidebands are created by strong electro-optic phase modulation at modulation frequency $f_m$ on the cw-laser. A programmable line-by-line waveshaper is then used to select a pair of comb lines as the local oscillators (see Fig. \ref{fig:data1}).  A periodic ramp voltage is applied on a phase modulator (PM) to scan the phase of the LOs. The LOs and the resonator pump laser are coherent with each other since they are derived from the same cw-laser. A detailed description of the experimental setup is provided in the Methods section.

The quadrature noise variances of 20 sets of comb pairs (40 qumodes) are measured by means of balanced homodyne detection. To measure the quadrature noise variance of qumodes $(-N,N)$, the EOM comb modulation frequency $f_m$ and the programmable waveshaper are adjusted to precisely match the frequencies of LO pairs to $\omega_p \pm N \times D_1$, where $N$ is the relative mode number from the mode being pump ($N = 0$), and $D_1/2\pi =21.95258$ GHz is the FSR of the resonator at 1551.5 nm wavelength. In each measurement, the phase of the LOs is ramped to yield varying quadrature variances. 
Figure \ref{fig:data1}(b) shows a representative quadrature noise variance (blue) relative to the shot noise (red) for qumodes (-4,4). A 30-point moving average is used to smooth out the phase fluctuations of the LOs. Raw squeezing of 1.6 $\pm$ 0.2 dB and anti-squeezing of 5.5 $\pm$ 0.1 dB are directly observed, and they are obtained by averaging the displayed extrema. The quadrature noise variances of all 40 qumodes are shown in Fig. \ref{fig:data1}(c), and squeezing/anti-squeezing are observed for all 40 qumodes. The number of measurable qumodes are limited by the 1 THz optical span of the EOM comb. All measurements are taken at 2.7 MHz frequency, 100 kHz resolution bandwidth, and 100 Hz video bandwidth on an electrical spectrum analyzer (ESA). The noise levels of qumodes (-1,1) to (-3,3) are not presented here as their measurements are affected by the transmitted amplified spontaneous emission (ASE) noise from the erbium-doped fiber amplifier (EDFA) near the pump frequency. This can be addressed in the future by increasing the intrinsic quality factor of the cavity and reducing the parametric oscillation threshold to eliminate the need for the EDFA. Finally, as shown in Fig. \ref{fig:data1}(d), no quantum correlation (two-mode squeezing) is observed for uncorrelated comb pairs. This serves as a critical check for our two-tone homodyne detection. 

The raw squeezing and anti-squeezing level of all 40 qumodes are summarized in Fig. \ref{fig:data2}(a). The raw squeezing in our experiment is primarily limited by the 83\% cavity escape efficiency, 1.7 dB optical loss and approximately 89\% photodiode quantum efficiency. The total efficiency after tapered fiber is 60$\%$.
Anti-squeezing levels near qumodes (-10,10), and from (-17,17) to (-23,23) are observed to be smaller than that of other qumodes. We suspect this is caused by spatial-mode interaction between different transverse mode families in the resonator. To characterize the frequency spectrum of the resonator, the relative mode frequencies, $\Delta \omega_N = \omega_N - \omega_0 - N \times D_1$, are measured with sideband spectroscopy method\cite{li2012sideband} and presented in Fig. \ref{fig:data2}(b), where $\omega_N$ is the resonance frequency of relative mode number $N$. An avoided mode crossing\cite{herr2014mode} was found near mode -8, and resonance frequencies below mode -18 and above mode 19 are observed to change abruptly. These are likely caused by the spatial-mode interaction and hybridization between two transverse cavity modes, which could degrade squeezing and anti-squeezing. This can be completely eliminated in the future by using a microresonator with a single transverse mode family \cite{kordts2016higher,yang2018bridging}. 


A qumode spectroscopy method is developed to characterize the frequency equidistance of squeezed qumodes, a prerequisite of frequency combs. Similar to classical cavity mode spectrum, we can define the relative qumode spectrum as $\Delta \omega^Q_N = \omega^Q_N - \omega_0 - N \times D_1$, where $\omega^Q_N$ is the optical frequency center of the $N$-th qumode. The relative qumode spectrum represents the qumode frequency deviation from equidistance. In order to identify the relative qumode spectrum, the two-sided squeezing/anti-squeezing spectral line shape is measured for each pair of qumodes, and the center frequency of the spectral line shape yields the relative qumode frequency. In the measurement, the $\pm N$-th LO frequencies are detuned by $\pm \delta$ from the equidistant frequencies, $\pm N \times$FSR, and noise variances are measured at each detuning point for qumodes (-$N$,$N$). For each pair of qumodes, the detuning ($\delta$) is varied from -30 MHz to + 30 MHz with an interval of 5 MHz, which sets the resolution of the line shape measurement. Measurements of qumodes (-4,4) at $\delta = -20,-10,0,10,20$ MHz are shown as examples in Fig. \ref{fig:data3}(b). At each detuning point, squeezing and anti-squeezing levels can be extracted by averaging the extrema. We plot the squeezing/anti-squeezing levels versus detuning ($\delta$) for all qumodes in Fig. \ref{fig:data3}(c), which manifest the two-sided spectral line shape of the qumodes. The squeezing/anti-squeezing extraction below 0.5 dB has relatively poor accuracy, but this does not affect the overall qumode spectrum envelopes.

The relative frequencies of the qumodes, i.e., relative qumode spectrum, can be obtained by identifying the centers of the anti-squeezing line shapes via Lorentzian fitting. The average root mean square deviation of the fitting is only 0.15 dB, showing an excellent agreement between fitting and measurements. $\Delta \omega^Q_N$ of all the qumodes are plotted in Fig. \ref{fig:data3}(d), and their deviations from equidistant are within the 5 MHz spectroscopy resolution limit for the entire 1 THz optical span of the quantum microcomb. The qumode spectrum overlaps well with the two-sided averaged cold cavity mode spectrum, $-(\Delta \omega_N + \Delta \omega_{-N})/2$, which represents the averaged deviation from equidistant of the cold cavity mode $N$ and $-N$. It should be noted that in the qumode spectrum measurement, the cavity is pumped by $> 100$ mW power, which could alter the cavity mode spectrum through thermo-optic effect and self/cross phase modulation effects. Further study in the future is necessary to understand the requirement for perfectly equidistant frequencies of qumodes. In this measurement, the cavity escape efficiency is adjusted to 77\% to achieve a stabler coupling condition as the entire measurement spans over 18 hours. As a result, the amount of squeezing/anti-squeezing at $\delta = 0$ MHz is different from that in the Fig. \ref{fig:data1}. 

In summary, we have demonstrated a deterministic two-mode-squeezed quantum microcomb on a silicon chip. The raw squeezing can be improved in the future by reducing system losses, improving photodiode quantum efficiency, and achieving higher resonator-waveguide escape efficiency. The number of measurable qumodes, 40, is primarily limited by the span of local oscillator, and this could be dramatically increased in the future by spectrum broadening of the EOM comb \cite{carlson2018ultrafast}, or by using broadband dissipative Kerr soliton microcombs\cite{kippenberg2018dissipative} as the local oscillators. $>$ 1,000 qumodes generation in a microresonator is possible when the dispersion is properly engineered. The miniaturization provides a path towards mass production of deterministic quantum frequency combs, which could be critical for applications in quantum computing, quantum metrology, and quantum sensing\cite{li2018quantum,lawrie2019quantum}.





\medskip

\noindent\textbf{Methods}
\begin{footnotesize}

\noindent{\bf Experimental setup.} 
The experimental setup is shown in Extended Data Fig. \ref{fig:setup}. A continuous-wave (cw) laser at 1551.5 nm is used to drive both the squeezed microcomb, and the local oscillators (LOs) for balanced homodyne detection. For the squeezed microcomb generation, the cw laser is amplified by an erbium-doped fiber amplifier (EDFA) to pump the Kerr microresonator. A fiber-Bragg grating (FBG) filter is used to filter out the amplified spontaneous emission (ASE) noise from the EDFA. The amplified pump laser is then coupled into the microresonator through a single mode tapered fiber. At the through port of the tapered fiber, another FBG filter is used to separate light at the pump laser wavelength from light at all other wavelengths. The transmitted squeezed microcomb from the FBG is sent to a 50/50 fiber coupler to be combined with the local oscillators for balanced homodyne detection. In the experiment, the Pound–Drever–Hall (PDH) locking technique is used to lock the pump laser frequency to the resonator mode frequency. This is implemented by phase modulating the pump laser before the EDFA with an electro-optic phase modulator (PM), and then photodetecting the pump laser after the second FBG filter. The phase modulation frequency is set to 80 MHz, much higher than the resonator linewidth.

The local oscillators in this experiment are derived from an electro-optic modulation (EOM) frequency comb \cite{murata2000optical}. The EOM comb is convenient to create coherent local oscillators which are hundreds of GHz apart from the pump laser frequency. In our EOM comb, the cw laser is amplified by an EDFA to 200 mW and is phase modulated by three cascaded electro-optic phase modulators at frequency $f_m$. The modulators are driven by amplified electrical signals that are synchronized by electrical phase shifters (PSs). The output power of the electrical amplifiers (Amps) is $\sim$ 33 dBm. As the EOM comb and the microresonator share the same pump laser, the local oscillators derived from the EOM comb are inherently coherent with the squeezed microcomb. A typical EOM comb spectrum is shown in Extended Data Fig. \ref{fig:EOM} (blue line), and the cw pump laser spectrum (black) is also shown as a reference. The EOM comb is then sent to a programmable line-by-line waveshaper, which can control the amplitude and phase of each EOM comb line. To measure the noise variance of qumodes $(-N,N)$, the waveshaper is set to only pass the comb lines whose frequencies are $\pm N \times $FSR apart from the pump laser. As an example, the local oscillators for qumodes (-21,21) are shown in Extended Data Fig. \ref{fig:EOM} (red line). Finally, the LOs are amplified to $\sim$ 17 mW and are combined with the squeezed microcomb for balanced homodyne detection. 

\noindent {\bf Characterization of balanced photodiodes.} In the two-mode squeezing noise variance measurement, the balanced photodiodes are operated in the shot noise limited regime. This is verified by the linear relationship between the noise power of the balanced photodiodes and the optical input power, which is shown in Extended Data Fig. \ref{fig:shotnoise}(a). The measurement is done at 2.7 MHz with 100 kHz resolution bandwidth (RBW). The electrical spectra from the balanced photodiodes at different optical input powers are shown in Extended Data Fig. \ref{fig:shotnoise}(b). The resonance peaks in the dark noise are likely caused by the electrical circuits in the balanced photodiodes. At 16.6 mW input power, the electrical spectrum is relatively flat. The spectra roll off around 20 MHz. 

\end{footnotesize}

\medskip

{\noindent \bf Data availability.} The data that support the plots within this paper and other findings of this study are available from the corresponding author upon reasonable request.

\medskip

\noindent\textbf{Acknowledgement}

\noindent The authors thank A. Beling at UVA for the access of signal generator, Y. Shen and J. Campbell at UVA for assisting photodiode quantum efficiency calibration, and gratefully acknowledge National Science Foundation.

\bibliographystyle{naturemag}
\bibliography{ref}  

\setcounter{figure}{0}
\renewcommand{\figurename}{Extended Data FIG.}

\begin{figure*}[!hbt]
\captionsetup{singlelinecheck=off, justification = RaggedRight}
\includegraphics[width=17cm]{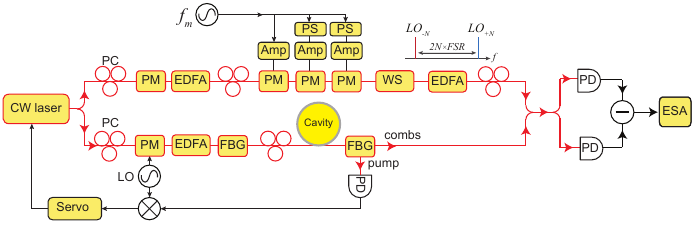}
\caption{{\bf Experimental setup.} Optical and microwave components are colored in red and black boxes, respectively. A continuous-wave (cw) laser drives both the squeezed microcomb and the local oscillators. Part of the cw laser is amplified by an erbium-doped fiber amplifier (EDFA) to pump the silica microresonator. A fiber-Bragg grating (FBG) filter is used at the microresonator through-port to separate the pump light and the squeezed light. The local oscillators are derived from an electro-optic modulation (EOM) frequency comb, which is driven by the same cw laser.
The cw laser is phase modulated by three tandem phase modulators (PMs) at frequency $f_{m}$. A programmable waveshaper (WS) is used to select a pair of comb lines to be the local oscillators. The LOs and the squeezed microcomb are combined and detected on the balanced photodetectors (PDs). The noise variance is characterized with an electrical spectrum analyzer (ESA). Polarization controller (PC), electrical amplifier (Amp), and phase shifter (PS) are also included in this figure.}
\label{fig:setup}
\end{figure*}

\begin{figure*}[!hbt]
\captionsetup{singlelinecheck=off, justification = RaggedRight}
\includegraphics[width=12cm]{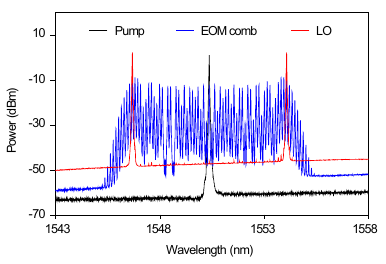}
\caption{{\bf Optical spectrum of local oscillators.} Optical spectra of the pump laser (black), the EOM frequency comb (blue), and the local oscillators (red) for qumodes (-21, 21).}
\label{fig:EOM}
\end{figure*}

\begin{figure*}[!hbt]
\captionsetup{singlelinecheck=off, justification = RaggedRight}
\includegraphics[width=17cm]{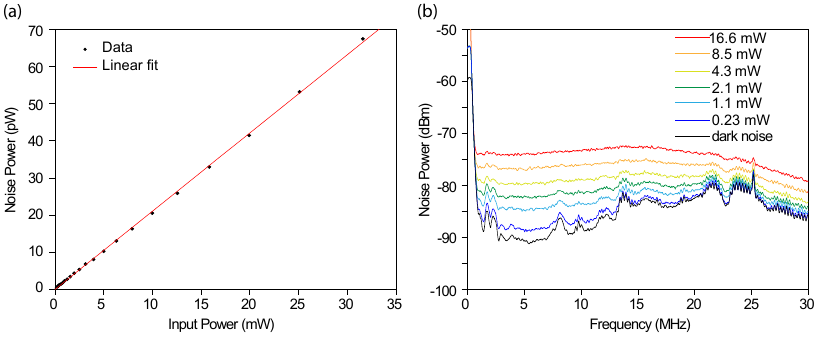}
\caption{{\bf {Noise characterization of the balanced photodiodes.}} {\bf{(a)}} Noise power vs. the optical power of local oscillators sent into the PDs. The noise power is measured at 2.7 MHz frequency, and the dark noise from the PDs has been subtracted from the noise power. The linear trend indicates the balanced photodiodes are operated in the  shot noise-limited regime. {\bf{(b)}} Electrical spectra of the balanced PD outputs at different local oscillator powers. All measurements in this figure are taken at 100 kHz resolution bandwidth.}
\label{fig:shotnoise}
\end{figure*}

\end{document}